\documentclass[a4paper,aps,pre,amsmath,amsfonts,showpacs,preprint]{revtex4}
\usepackage{graphicx}
\usepackage{amssymb}

\begin{document}

\title{Inhomogeneous sandpile model: Crossover from multifractal scaling
to finite size scaling }

\author{Jozef \v{C}ern\'{a}k }

\email{jcernak@kosice.upjs.sk}

\affiliation{University of P. J. \v{S}af\'{a}rik in Ko\v{s}ice, Department
of Biophysics, Jesenn\'{a} 5, SK-04000 Ko\v{s}ice, Slovak Republic}

\date{\today{}}

\begin{abstract}
We study an inhomogeneous sandpile model in which two different toppling
rules are defined. For any site only one rule is applied corresponding
to either the Bak, Tang and Wiesenfeld model {[}P.Bak, C. Tang, and K. Wiesenfeld, 
Phys. Rev. Lett. \textbf{59}, 381 (1987){]} or the Manna two-state sandpile model 
{[}S. S. Manna, J. Phys. A \textbf{24}, L363 (1991){]}. A parameter $c$ is introduced 
which describes a density of sites which are randomly deployed and where
the stochastic Manna rules are applied. The results show that the
avalanche area exponent $\tau_{a}$, avalanche size exponent $\tau_{s}$,
and capacity fractal dimension $D_{s}$ depend on the density $c$.
A crossover from multifractal scaling of the Bak, Tang, and Wiesenfeld
model ($c=0$) to finite size scaling was found. The critical density
$c$ is found to be in the interval $0<c<0.01$. These results demonstrate
that local dynamical rules are important and can change the global
properties of the model.
\end{abstract}


\maketitle

\section{Introduction}

Bak, Tang, and Wiesenfeld (BTW) \cite{BTW_1987} introduced a concept
of self-organized criticality (SOC) as a common feature of different
dynamical systems where the power-law temporal or spatial correlations
are extended over several decades. Dynamical systems with many interacting
degrees of freedom and with short range couplings naturally evolve
into a critical state through a self-organized process. They proposed
a simple cellular automaton with deterministic rules, which is known
as a sandpile model, to demonstrate this new phenomenon. In this model
the relaxation rules are conservative, no dissipation takes place
during relaxation, and correspond to a nonlinear diffusion equation
\cite{BTW_1987}. Generally, the sandpile model is represented by
a $d$-dimensional hypercube of the finite linear size $L$. Its boundaries
are open and allow an energy dissipation, which takes place only at
the boundaries. 

Manna proposed a two-state version of the sandpile model \cite{Manna_1991}
where no more than one particle is allowed to be at a site in the
stationary state. If one particle is added to a randomly chosen site,
then relaxation starts depending on the occupancy of the site. If
the site is empty, a particle is launched. In the case when the site
is not empty, a hard core interaction throws the particles out from
the site and the particles are redistributed in a random manner among
its neighbours. All sites affected by this redistribution create an
avalanche. An avalanche is stopped if any site reached the stationary
state, i.e. no more than one particle occupies a site.

The first systematic study of scaling properties, universality and
classification of deterministic sandpile models was carried out by
Kadanoff \emph{et al.} \cite{Kadanoff}. Using numerical simulations
and by varying the underlying microscopic rules which describe how
an avalanche is generated they investigated whether different models
have the same universal properties. Applying finite-size scaling (FSS)
and multifractal scaling techniques they studied how a finite-size
of the system affects scaling properties. 

The real-space renormalization group calculations \cite{Pietronero}
suggested that deterministic \cite{BTW_1987} and stochastic \cite{Manna_1991}
sandpile models belong to the same universality class. On the other
hand, many numerical results \cite{Ben-Hur,Biham,Lubeck,Menech,Milsh}
show clearly two different universality classes. They do not confirm
the hypothesis that small modifications in the dynamical rules of
the models do not change the universality class, presented by Chessa
\emph{at al}. \cite{Chessa}. 

This study was motivated by the results published by Tebaldi \emph{et
al.} \cite{Tebaldi_1999}, and Stella and Menech \cite{Stella_2001},
where a multifractal scaling of an avalanche size distribution of
the BTW model was demonstrated. They assume that a multifractal character
for SOC models like the BTW model is a crucial step towards the solution
of universality issues. By applying the moment analysis they found
FSS for the two-state Manna model \cite{Stella_2001}. Based on these
results they conclude that the 2D BTW model and the Manna model belong
to qualitatively different universality classes. This assumption was
confirmed recently \cite{Karma,Karma_E}, where a precise toppling
balance has been investigated in more detail.

In this paper we report the results of disturbing the dynamics of
the BTW model using stochastic Manna sites which are randomly deployed.
They can introduce stochastic events during an avalanche propagation.
Our model was derived from the inhomogeneous sandpile model \cite{Cer}
in witch two different deterministic toppling rules were defined.
In the proposed model the first toppling rule corresponds to the BTW
model \cite{BTW_1987} and the second rule is now stochastic and corresponds
to the two-state Manna model \cite{Manna_1991}. The model is similar
to that in Ref. \cite{Karma_E}, however we applied the original toppling
rules of the listed sandpile models.

The paper is organized as follows. The inhomogeneous sandpile model
is introduced in Sec. \ref{sec:Mathematical-model}. The avalanche
scaling exponents, capacity fractal dimensions and crossover from
multifractal to FSS are investigated with numerical simulations and
the results are presented in Sec. \ref{sec:Results}. The Sec. \ref{sec:Discussion}
is devoted to a discussion which is followed by conclusions in Sec.
\ref{sec:Conclusion}.

\section{\label{sec:Mathematical-model}Mathematical model}

We consider a $d$-dimensional hyper-cubic lattice of linear size
$L$, and a notation presented by Ben-Hur \emph{et al.} \cite{Ben-Hur}
is followed to define a sandpile model. Each site $\mathbf{\mathbf{i}}$
has assigned a dynamical variable $E(\mathbf{\mathbf{i}})$ that generally
represents a physical quantity such as energy, grain density, stress,
etc. A configuration $\{ E(\mathbf{i})\}$ is classified as stable
if for all sites $E(\mathbf{\mathbf{i}})<E_{c}$, where $E_{c}$ is
a threshold value. We note that the two-state Manna model \cite{Manna_1991}
has no threshold $E_{c}$. The Manna model has defined a hard core
repulsion interaction among different particles at the same position.
This hard core repulsion interaction can be described by a threshold
where the threshold value $E_{c}=2$ is assigned to any site. In our
inhomogeneous sandpile model, the threshold values $E_{c}$ depend
on the site position $\mathbf{i}$, $E_{c}(\mathbf{i})$ \cite{Cer}.
The conditions for a stationary state, a stable configuration $\{ E(\mathbf{i})\}$
(no avalanche), are now $E(\mathbf{i})<E_{c}(\mathbf{i})$, where
the threshold $E_{c}(\mathbf{i})$ at the site $\mathbf{i}$ was randomly
chosen from two allowed values 

\begin{equation}\label{eq:thresholds} E_{c}(\mathbf{i})=\begin{cases} E_{c}^{I}\ =4\\

E_{c}^{II}=2. \end{cases}\end{equation}For any site $\mathbf{i}$ the threshold $E_{c}(\mathbf{i})$ [Eq.
(\ref{eq:thresholds})] is defined in such a manner that $n$ randomly
chosen sites have the value $E_{c}^{II}$ and the remaining $L^{d}-n$
sites have the value $E_{c}^{I}$. The density of sites with the threshold
value $E_{c}^{II}$ is denoted $c$, and $c=n/L^{d}$.

Let us assume that a stable configuration $\{ E(\mathbf{j})\}$ is
given, and then we select a site $\mathbf{i}$ at random and increase
$E(\mathbf{i})$ by some amount $\delta E$. We now consider $\delta E=1$
for any site. When an unstable configuration is reached, $E(\mathbf{i})\geq E_{c}(\mathbf{i})$,
a relaxation takes place. An unstable site $\mathbf{i}$ lowers its
energy, that is distributed among the neighbor sites. The directions
to the neighbor sites are defined by the vectors $\mathbf{e}_{1}=(0,1)$,
$\mathbf{e}_{2}=(0,-1)$, $\mathbf{e}_{3}=(1,0)$, and $\mathbf{e}_{4}=(-1,0)$.
The relaxation is defined by the following rules

\begin{equation}
E(\mathbf{i})\rightarrow E(\mathbf{i})-\sum_{e}\Delta E(\mathbf{i}),\label{eq:relax1}\end{equation}

\begin{equation}
E(\mathbf{i}+\mathbf{e})\rightarrow E(\mathbf{i}+\mathbf{e})+\Delta E(\mathbf{e}),\label{eq:relax2}\end{equation}

\begin{equation}
\sum_{e}\Delta E(\mathbf{e})=E_{c}(\mathbf{i}),\label{eq:relax3}\end{equation}

\begin{equation}\label{eq:relax4} \mathbf{e}=\begin{cases} \{\mathbf{e}_{1},\  \mathbf{e}_{2},\ \mathbf{e}_{3},\ \mathbf{e}_{4}\} & if\ E_{c}(\mathbf{i})=E_{c}^{I}\\ 

\{\mathbf{e}_{\zeta},\ \mathbf{e}_{\eta}\} & if\ E_{c}(\mathbf{i})=E_{c}^{II}\end{cases} \end{equation}where $\mathbf{e}$ is a set of vectors from the site $\mathbf{i}$
to its neighbors. The indexes $\zeta$ and $\eta$ are integers $1,2,3$,
and $4$ randomly chosen at any relaxation. The neighbors that receive
the energy can became unstable and topple, thus generating an avalanche.
The distribution of energy is described by Eqs. (\ref{eq:relax1})
and (\ref{eq:relax2}), we added additional rules Eqs. (\ref{eq:relax3})
and (\ref{eq:relax4}) which specify the manner how the energy is
distributed depending on the position $\mathbf{i}$, threshold $E_{c}(\mathbf{i})$
[Eq. (\ref{eq:relax3})], and corresponding sandpile model [Eq. (\ref{eq:relax4})].
The relaxation rules Eqs. (\ref{eq:relax1})-(\ref{eq:relax4}) are
applied until that moment when a new stable configuration is reached
again, for all sites $E(\mathbf{i})<E_{c}(\mathbf{i})$. Obviously,
during one avalanche an arbitrary unstable site $\mathbf{i}$ can
transfer the energy $E_{c}(\mathbf{i})$ a few times to became stable,
$E(\mathbf{i})<E_{c}(\mathbf{i})$. A $d$-dimensional lattice has
open boundaries so added energy can flow outside the system, and an
energy dissipation takes place only at the boundaries. 

This model has been designed to enable a well defined change between
two well known nondirected sandpile models: deterministic \cite{BTW_1987}
and stochastic \cite{Manna_1991} (nondirected only on average) similarly
as in Ref. \cite{Karma}. The model belongs to the critical height
models with conservative relaxation rules and with undirected energy
transfer where the two thresholds are randomly frozen. It can be characterized
as a sandpile with a possibility to modify its scaling behaviors.

\section{\label{sec:Results}Results}

We shall report the results obtained using numerical simulation of
the conservative, undirected, critical height sandpile model defined
by Eqs. (\ref{eq:relax1})-(\ref{eq:relax4}). The simulations were
carried out for the following parameters: $d=2$, two-dimensional
lattice of linear sizes $L=256,\ 512$ and $1024$, randomly added
energy $\delta E=1$, two thresholds either $E_{c}^{I}=4$ or $E_{c}^{II}=2$,
and with density of sites with threshold $E_{c}^{II}$ in the interval
$0\leq c\leq1$. In our simulations we have used the density $c$
as a model parameter. For densities of stochastic sites $c=0$ and
$1$ the model behaves as the BTW model \cite{BTW_1987} and Manna
model \cite{Manna_1991}, respectively, which are both considered
to be \emph{Abelian} \cite{Dhar}. 

Avalanches can be characterized by such properties as their size,
area, lifetime, linear size, and perimeter. We concentrate only on
a minimal number of parameters which are necessary to demonstrate
the investigated phenomena: the avalanche area $a$ and avalanche
size $s$. Here the avalanche area $a$ is the number of lattice sites
that have relaxed at least once during the avalanche. The avalanche
size $s$ is the total number of relaxations that occurred during
the avalanche. The probability distributions of these variables are
usually described as power-laws with cutoff

\begin{equation}
P(x)=x^{-\tau_{x}}F(x/x_{c}),\label{eq:FSS}\end{equation}
where $x=a,s.$ When the system size $L$ goes to infinity, the cutoff
$x_{c}$ diverges as $x_{c}\sim L^{D_{x}}$. If we assume FSS, then
the set of exponents ($\tau_{x},D_{x}$) from Eq. (\ref{eq:FSS})
defines the universality class of the model \cite{Chessa}.

The avalanche area probability distribution $P(a)$ and avalanche
size probability distributions $P(s)$ have been analyzed at finite
lattice sizes $L=256,512$, and $1024$. It is expected that these
distributions follow a power-law $P(x)\sim x^{\tau_{x}}$ [Eq. (\ref{eq:FSS})].
For any lattice size $L$ and density $c$ the corresponding scaling
exponents $\tau_{x,L}(c)$ were determined. The scaling exponents
found in the numerical simulations for the largest lattice size $L=1024$
and for selected densities $c$ are presented in Table \ref{cap:The-scaling-exponents}.
It is evident that the exponents are increasing with $c$ in the interval
$0<c<0.1$ and then for densities $c>0.1$ they are almost constant.

\begin{table}[b]

\caption{\label{cap:The-scaling-exponents}The scaling exponents $\tau_{x,L=1024}(c)$
for the finite lattice size $L=1024$ and selected densities in the
interval $0\leq c\leq1$. The statistical errors are $\pm0.001$.}

\begin{tabular}{lccccc}
\hline 
density $c$&
$0$&
$0.01$&
$0.10$&
$0.50$&
$1$\tabularnewline
\hline
$\tau_{a,L=1024}$&
$1.131$&
$-$&
$1.291$&
$1.315$&
$1.338$\tabularnewline
$\tau_{s,L=1024}$&
$1.137$&
$1.240$&
$1.263$&
$1.266$&
$1.283$\tabularnewline
\hline
\end{tabular}
\end{table}

The scaling exponents $\tau_{x,L}$ show a finite size-effect when
the lattice size $L$ is changed. Their dependences on lattice sizes
$L$ are approximated by a formula proposed by Manna \cite{Manna_PA}

\begin{equation}
x=x_{L\rightarrow\infty}-\frac{const.}{\ln(L)}.\label{eq:fit}\end{equation}
 This approximation was used to extrapolate the scaling exponents
$\tau_{x,L\rightarrow\infty}$ for the infinite lattice $L\rightarrow\infty$. 

\begin{figure}[t]
\includegraphics{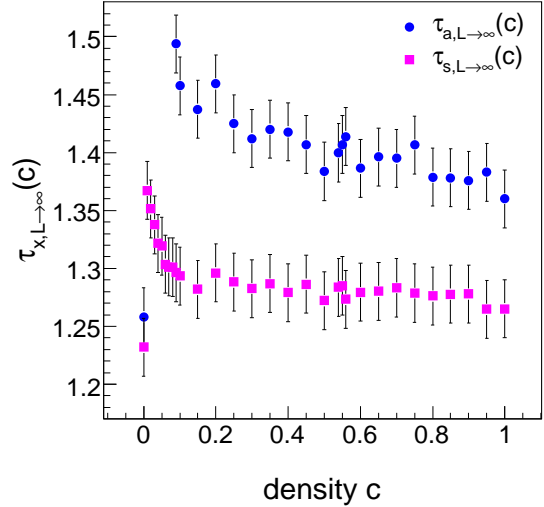}

\caption{\label{cap:exponents}(Color online) The avalanche area and size
scaling exponents $\tau_{a,L\rightarrow\infty}$ and $\tau_{s,L\rightarrow\infty}$
are approximated for the infinite lattice size $L\rightarrow\infty$.
The exponents depend on the density $c$ of Manna sites. }
\end{figure}

The avalanche size probability distributions $P(s)$ obey the power-law
dependence for any density $c$. The corresponding scaling exponents
$\tau_{s,L\rightarrow\infty}(c)$ are shown in the Fig. \ref{cap:exponents}.
In the range of densities $0.01\leq c\leq0.1$ these scaling exponents
decrease from $\tau_{s,L\rightarrow\infty}(0.01)=1.37$$\pm0.025$
to $\tau_{s,L\rightarrow\infty}(0.1)=1.29\pm0.025$ and then, for
higher densities $c>0.1$, are almost constant. 

\begin{figure}[t]
\includegraphics{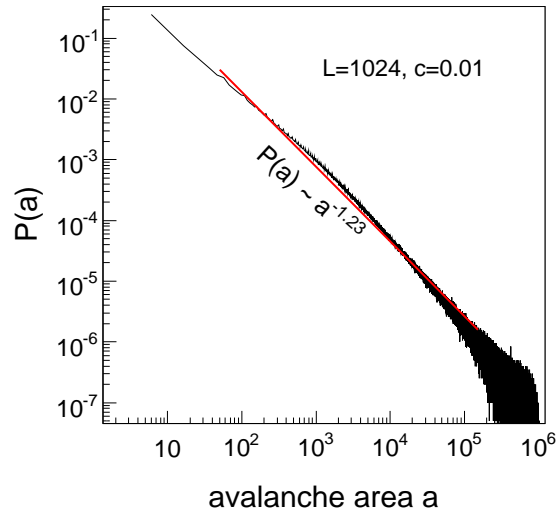}

\caption{\label{cap:avalanche-area}(Color online) The avalanche area distribution
$P(a)$ does not follow exactly a power-law function. The parameters
used in the numerical simulation were: density $c=0.01$ and linear
lattice size $L=1024$. }
\end{figure}

The avalanche area scaling exponents $\tau_{a,L\rightarrow\infty}$
show a more complex dependence on the density $c$. For densities
$0.09\leq c\leq0.5$ they decrease from $\tau_{a,L\rightarrow\infty}(0.09)=1.49$$\pm0.025$
to $\tau_{a,L\rightarrow\infty}(0.5)=1.38\pm0.025$, then for higher
densities $c>0.5$ the exponents $\tau_{a,L\rightarrow\infty}(c)$
are almost constant. It was observed that for densities $0.01\leq c\leq0.09$
the avalanche area distributions $P(a)$ do not follow exactly a power-law
dependence as it is expected from Eq. (\ref{eq:FSS}). Therefore the
exponents $\tau_{a,L\rightarrow\infty}(c)$ from this density interval
are not included in Fig. \ref{cap:exponents}. One typical example
is shown in Fig. \ref{cap:avalanche-area} where the density of random
toppling sites is $c=0.01$ and the lattice size is $L=1024$. The
double-log plot of area distribution function $P(a)$ clearly shows
that a possible approximation function is not a straight line which
must correspond to the simple power-law dependence. 

For the two well known sandpile models, BTW ($c=0$) and Manna ($c=1$)
the scaling exponents $\tau_{a,L\rightarrow\infty}(0)=1.26$, $\tau_{s,L\rightarrow\infty}(0)=1.23$,
$\tau_{a,L\rightarrow\infty}(1)=1.36$, and $\tau_{s,L\rightarrow\infty}(1)=1.27$
were found. In addition, for all densities $c$ (see Fig. \ref{cap:exponents})
the relation $\tau_{a,L\rightarrow\infty}(c)>\tau_{s,L\rightarrow\infty}(c)$
is valid. 

\begin{figure}[t]
\includegraphics{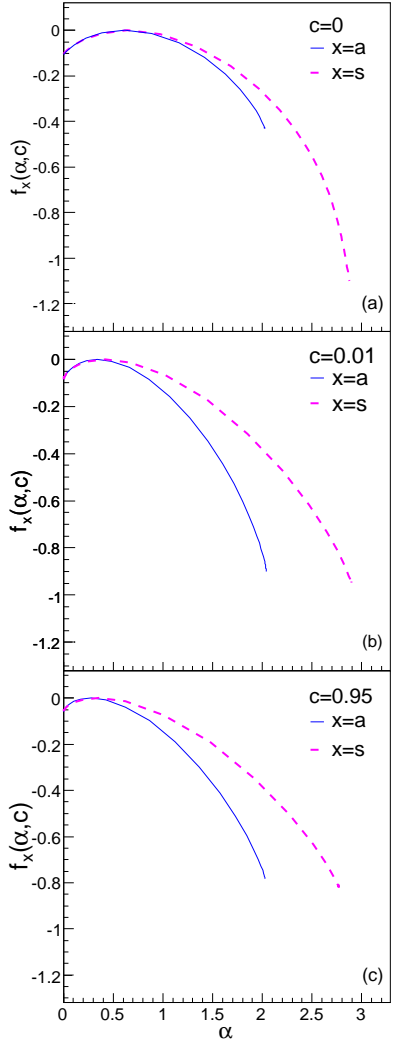}

\caption{\label{cap:multi}(Color online) The extrapolated spectra of the
avalanche area $a$ and avalanche size $s$ for various densities
of Manna toppling sites: (a) $c=0$ BTW model, which shows multifractal
scaling, (b) $c=0.01$ at which the multifractal scaling of BTW model
is destroyed, and (c) $c=0.95$ where the model shows the FSS near
the two-state Manna model ($c=1$). The maximal error bars of $f(\alpha,c)$
are for $q\approx0$, and are approximately $\pm0.05$, but for a
higher $q$ they are smaller. The $\alpha$ values are determined
within errors $\pm0.025$. }
\end{figure}

The scaling exponents $\tau_{x,L}$ as functions of the lattice size
$L$ show a finite-size scaling effect [Eq. (\ref{eq:fit})]. An exact
determination of scaling exponents $\tau_{x,L\rightarrow\infty}$
from numerical experiments is therefore a difficult task. A new method
was introduced \cite{Lubeck} to increase the numerical accuracy of
the exponents based on their direct determination. We found that the
method gives slightly larger exponents than a simple extrapolation
of Eq. (\ref{eq:fit}). However, the exponents $\tau_{s}$ do not
fluctuate around their mean values as it was observed in the paper
\cite{Lubeck}. Our error bars were larger, therefore we have to repeat
this analysis again in more details. 

Tebaldi \emph{et al.} \cite{Tebaldi_1999} found that in the BTW model
the avalanche area distributions $P(a)$ show FSS and avalanche size
distribution $P(s)$ scale as a multifractal. To describe these scaling
properties rather a multifractal spectrum $f(\alpha)$ versus $\alpha$
than the single scaling exponent $\tau_{s}$ [Eq. (\ref{eq:FSS})]
is necessary. Thus, the scaling exponent $\tau_{s}$ loses the importance
and is replaced by a spectrum of exponents. Despite this fact, the
avalanche size scaling exponents $\tau_{s,L\rightarrow\infty}(0)$
are determined. They enable a comparison with the previous results,
since the whole point is that the exponent $\tau_{s,L\rightarrow\infty}(0)$
does not exist. The recent studies \cite{Tebaldi_1999,Stella_2001}
led us to analyze the multifractal properties of the model given by
Eqs. (\ref{eq:relax1})-(\ref{eq:relax4}) for various densities $c$.
To determine the multifractal spectra a method presented in the paper
\cite{Stella_2001} was useful. There, for any finite-size lattices
$L$, the quantities $\alpha_{x}(q,L)=\left\langle log(x)x^{q}\right\rangle /\left[log(L)\left\langle x^{q}\right\rangle \right]$
and $\sigma_{x}(q,L)\sim log\left(\left\langle x^{q}\right\rangle \right)/log(L)$
were computed. It was observed that $\alpha_{x}(q,L)$ and $\sigma_{x}(q,L)$
show a finite-size dependence on the system size $L$, which is well
approximated by Eq. (\ref{eq:fit}) and this relation was used to
extrapolate $L\rightarrow\infty$ quantities. Based on the Legendre
structure relating $f_{x}$ to $\sigma_{x}$, a parametric representation
of $f_{x}(\alpha_{x})$ by plotting $f_{x}(q)=\sigma_{x}(q)-\alpha_{x}(q)q$
versus $\alpha_{x}(q)$ can be obtained \cite{Stella_2001}. 

Some significant spectra of $f_{x}(\alpha_{x},c)$ extrapolated for
an infinite lattice size $L\rightarrow\infty$ are shown for illustration
in Fig. \ref{cap:multi}. The $f_{x}(\alpha_{x},c)$ values were determined
for the parameter $q$ in the range $-3.5<q<3.5$ and they are limited
by errors about $\pm0.08$, similarly as in Ref. \cite{Stella_2001}.
We have observed that if $f_{x}(\alpha_{x},c)$ spectra are computed
for all avalanches where $a>50$ then the errors of $f_{x}(\alpha_{x},c)$
are $\pm0.05$. The multifractal scaling of the avalanche size probability
distribution $P(s)$ and FSS of avalanche area probability distribution
$P(a)$ were found at density $c=0$ (see Fig. \ref{cap:multi} (a)).
The avalanche probability distributions $P(x)$ show FSS for densities
$c=0.01$ Fig. \ref{cap:multi} (b), and for $c=0.95$ Fig. \ref{cap:multi}
(c) which is close to the Manna model ($c=1$). The spectra for $c=0$
and $1$ agree well with the previous results \cite{Stella_2001}.
It was found that the multifractal scaling of $P(s)$ was destroyed
(Fig. \ref{cap:multi}(b)) at a relatively small density of Manna
sites $0<c<0.01$. %
\begin{figure}[t]
\includegraphics{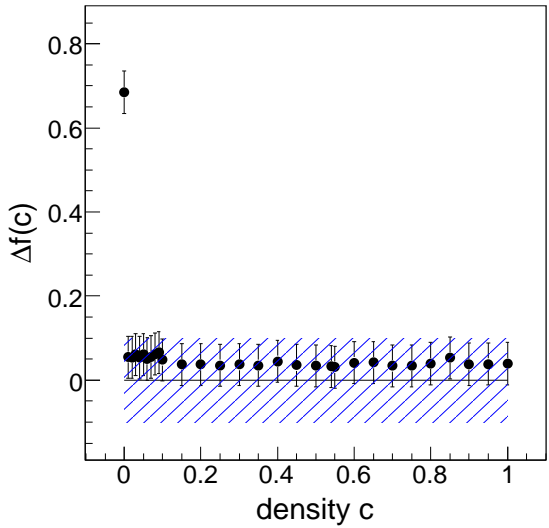}

\caption{\label{cap:-crossover}(Color online) A crossover from multifractal
scaling to finite size scaling takes place at $0<c<0.01$. The hatched
area border an interval of $\Delta f(c)$ in which $\Delta f(c)\doteq0$
and probability distributions $P(x)$ show finite size scaling. We
note that $\Delta f(c)=f_{a}^{min}(\alpha_{a},c)-f_{s}^{min}(\alpha_{s},c)$. }
\end{figure}

Stella \emph{et al.} \cite{Stella_2001} claim that if probability
distributions $P(x)$ satisfied FSS the large $q$ data accumulate
in the same value $f_{x}(\alpha_{x})$ where $\alpha_{x}^{max}=D_{x}$
and $f_{x}=-(\tau_{x}-1)D_{x}$. However, for probability distribution
showing the multifractal scaling there is no accumulation point and
$f_{x}(\alpha_{x})$ points shift progressively down as the parameter
$q$ is increasing and the parameter $q$ approaches $D_{x}$. This
fact is utilized as a simple criterion to recognize which probability
distributions show either multifractal scaling or FSS \cite{Stella_2001}.
The equality $f_{a}^{min}(\alpha_{a},c)\doteq f_{s}^{min}(\alpha_{s},c)$
is considered to be an attribute that probability distributions $P(x)$
show FSS. To test this equality the differences $\Delta f(c)$ defined
as $\Delta f(c)=f_{a}^{min}(\alpha_{a},c)-f_{s}^{min}(\alpha_{s},c)$
were determined. The equality $\Delta f(c)\doteq0$ is considered
for true if $\left|\Delta f(c)\right|\leq0.10$ which reflects numerical
errors. The differences $\Delta f(c)$ are shown in Fig. \ref{cap:-crossover}
where the hatched area limits the region where the equality is true
and thus the avalanche probability distributions $P(x)$ show FSS
behavior. It is clearly evident that only one value of $\Delta f(c)$
at the density $c=0$, is outside the region $\left|\Delta f(0)\right|>0.10$,
and it corresponds to multifractal scaling of the BTW model \cite{Tebaldi_1999,Stella_2001}.
We have no data from the interval of densities $0<c<0.01$ and thus
we may only expect that a crossover from multifractal to FSS takes
place in this interval. 

\begin{figure}[b]
\includegraphics{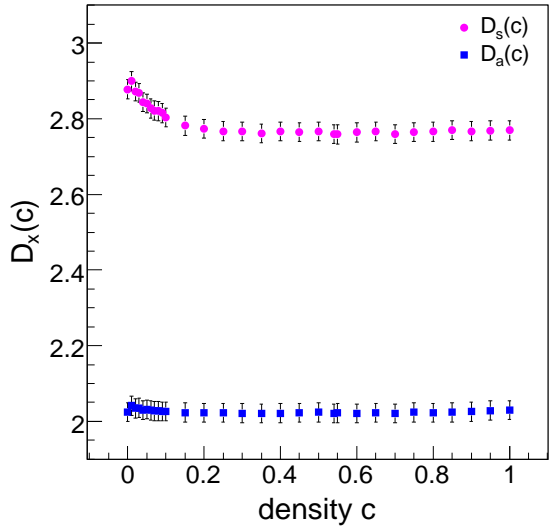}

\caption{\label{cap:capacity}(Color online) The capacity fractal dimensions
$D_{x=a,s}(c)$ as functions of the density $c$. The error bars are
$\pm0.025$. }
\end{figure}

The $f_{x}(\alpha_{x},c)$ spectra enable us to determine the capacity
fractal dimensions $D_{x}(c)$ as $D_{x}(c)=\alpha_{x}^{max}(c)$.
The results $D_{x}(c)$ for densities $0\leq c\leq1$ are shown in
the Fig. \ref{cap:capacity}. For the BTW model $D_{s}(0)=2.88\pm0.025$
and $D_{a}(0)=2.02\pm0.025$, and for the Manna model $D_{s}(1)=2.77\pm0.025$
and $D_{a}(1)=2.03\pm0.025$ were found. The avalanche area capacity
fractal dimensions $D_{a}(c)$ are almost constant $D_{a}(c)\doteq2$,
for any density $c$, and $D_{a}(0)\doteq D_{a}(1)$. In the interval
of densities $0.01<c<0.15$ the avalanche size dimension $D_{s}(c)$
is decreasing from $D_{s}(0.01)=2.90$ to the value $D_{s}(0.15)=2.78$
and is then almost constant for $c>0.15$, finally $D_{s}(0)>D_{s}(1)$.

\begin{figure}
\includegraphics{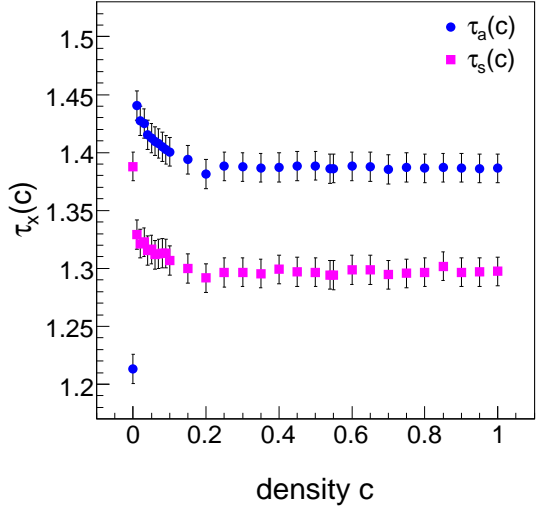}

\caption{\label{cap:f_exponets}(Color online) The scaling exponents were
determined using $\tau_{x}(c)=1-f_{x}^{min}(c)/D_{x}(c)$, for the
moment analysis all avalanches where $a>50$ were taken into account. }
\end{figure}

The moment analysis method \cite{Stella_2001} was used to clarify
interesting properties of the scaling exponents $\tau_{x,L\rightarrow\infty}(c)$
which are shown in Fig. \ref{cap:exponents}. The values of the functions
$f_{x}^{min}(c)$ and $D_{x}(c)$ ( Fig. \ref{cap:capacity}) are
determined from the $f_{x}(\alpha_{x},c)$ plots. For specific densities
$c=0$ (the BTW model) and $c=1$ (the Manna model) $f_{a}^{min}(0)=-0.43\pm0.05$
and $f_{s}^{min}(1)=-0.784\pm0.05$ were found. Then the scaling exponents
are given $\tau_{x}(c)=1-f_{x}^{min}(c)/D_{x}(c)$ and are shown in
the Fig. \ref{cap:f_exponets}. For the density $c=0$, it was found
$\tau_{a}(0)=1.213\pm0.0125$. For the densities $0.01\leq c\leq0.15$,
the exponents decrease from $\tau_{a}(0.01)=1.441\pm0.0125$ and $\tau_{s}(0.01)=1.329\pm0.0125$
to the values $\tau_{a}(0.15)=1.394\pm0.0125$ and $\tau_{s}(0.15)=1.299\pm0.0125$,
which are subsequently constant for $c>0.15$. For the density $c=1$,
they are $\tau_{a}(1)=1.386\pm0.0125$ and $\tau_{s}(1)=1.297\pm0.0125$.
These results are similar to those determined directly from the distribution
functions $P(x)\sim x^{-\tau_{x}}$(Fig. \ref{cap:exponents}). 

\section{\label{sec:Discussion}Discussion}

The plots of $\tau_{x,L}$ vs. $1/\ln L$ and an approximation given
by Eq. (\ref{eq:fit}) were used to extrapolate scaling exponents
$\tau_{x,L\rightarrow\infty}$ \cite{Manna_PA,Lubeck}. L\"{u}beck
and Usadel \cite{Lubeck} have analyzed an influence of an uncertainty
in the determination of the exponents $\tau_{x,L}$ on the precision
of the extrapolated exponents $\tau_{x,L\rightarrow\infty}$. Their
results show that this method is not very accurate.  However, this
approximation enables us to make a comparison of our results with
previous ones. The scaling exponents of the BTW model $\tau_{a,L\rightarrow\infty}(0)=1.26$
and $\tau_{s,L\rightarrow\infty}(0)=1.23$ (Fig. \ref{cap:exponents})
are approximately the same as those found in Ref. \cite{Lubeck} ($\tau_{a}=1.258$
and $\tau_{s}=1.247$) using the same method. The exponents of the
Manna model $\tau_{a,L\rightarrow\infty}(1)=1.36$, and $\tau_{s,L\rightarrow\infty}(1)=1.27$
are comparable with the previous results, $\tau_{s,L=1024}=1.28\pm0.02$
\cite{Manna_1991} and with $\tau_{a}=1.373$ and $\tau_{s}=1.275$,
which were found by direct determination of exponents \cite{Lubeck}
or calculated from the moment analysis $\tau_{a}\doteq1.36$ and $\tau_{s}\doteq1.28$
\cite{Lubeck_2}. The results obtained by the moment analysis \cite{Stella_2001},
$f_{a}^{min}(0)=-0.43\pm0.05$ and $f_{s}^{min}(1)=-0.784\pm0.05$,
agree well with the previous results, ${\textstyle \sigma_{a}=-0.391\pm0.011}$
and $\sigma_{s}=-0.7900\pm0.002$ \cite{Lubeck_2}. We may conclude
that the experimental data for two known densities, $c=0$ and $1$,
and data analysis methods give approximately the same exponents as
were found in previous numerical experiments \cite{Manna_1991,Lubeck,Lubeck_2}. 

The scaling exponents defined by Eq. (\ref{eq:FSS}) \cite{Kadanoff}
and the conditional exponents $\gamma_{xy}$\cite{Christensen,Ben-Hur}
can characterize the sandpile models. The theory predicts $\tau_{s}=1.253$
\cite{Pietronero} and a few numerical experiments show $D_{s}\simeq2.7$
and $D_{a}\simeq2$ \cite{Pietronero,Chessa}. The conditional exponents
$\gamma_{sa}$ determined directly from the numerical experiments
are $\gamma_{sa}(0)=1.06$ and $\gamma_{sa}(1)=1.23$ \cite{Ben-Hur}.

Let us assume that the BTW and Manna models belong to the same universality
class. Then the scaling exponents $\tau_{x}(c),D_{x}(c)$ [Eq. (\ref{eq:FSS}]
of the model (Eqs. (\ref{eq:relax1})-(\ref{eq:relax4})) must be
independent on the density $c$, i.e. $\tau_{x}(c)=const.$ and $D_{x}(c)=const.$
This means that knowing only the scaling exponents ($\tau_{x}(c),D_{x}(c)$),
we could not distinguish how many sites are toppling by deterministic
or stochastic manner [Eq. (\ref{eq:relax4})]. 

We observed that the capacity fractal dimensions $D_{a}(c)$ is constant
for any density $c$, $D_{a}(c)\doteq2$. The capacity fractal dimension
$D_{s}(0)=2.88$ is the same as was found in the Ref. \cite{Stella_2001},
$D_{s}\doteq2.86$ (determined from the Fig. 1(a) in \cite{Stella_2001}).
Our capacity fractal dimension $D_{s}(1)=2.77$ is higher than the
value $D\simeq2.7$ \cite{Manna_1991,Chessa}, however it is closer
to the $D\simeq2.75$ \cite{Karma}. In addition, for densities $0.01\leq c\leq0.1$,
the scaling exponents $\tau_{x,L\rightarrow\infty}(c)$, $\tau_{x}(c)$
(Figs. \ref{cap:exponents} and \ref{cap:f_exponets}) and $D_{s}(c)$
(Fig. \ref{cap:capacity}) depend on the density $c$. These scaling
exponents and capacity fractal dimension are not constant. They demonstrate
that the assumption about a single universality class is wrong and
thus confirm the existence of different universality classes. 

The conditional scaling exponents $\gamma_{xy}$ \cite{Christensen}
can be determined as $\gamma_{xy}(c)=(\tau_{y}(c)-1)/(\tau_{x}(c)-1)$
\cite{Lubeck_2}. Substituting the known scaling exponents $\tau_{x}(c)$
(Fig. \ref{cap:f_exponets}), we determined $\gamma_{sa}(0.01)\doteq1.34$
and for the Manna model, $\gamma_{sa}(1)\doteq1.29$. We note that
the scaling exponent $\tau_{s}(0)$ does not really exist. 

To determine the exact scaling exponents of the probability distribution
functions $P(x)$, the experimental data must show a power-law dependence
given by Eq. (\ref{eq:FSS}). However, the avalanche area size distributions
$P(a)$ do not follow exactly power-law distributions for densities
$0<c\leq0.1$ in the whole range of avalanche area sizes, a typical
example is shown in the Fig. \ref{cap:avalanche-area}. Chessa \emph{et
al.} \cite{Chessa} found that the area size distribution $P(a)$
of the BTW model ($c=0$) is not compatible with the FSS hypothesis
in the whole range of avalanches. However, for large size of avalanches
the FSS form must be approached. They assume that the scaling in the
BTW model needs sub-dominant corrections of the form $P(x)=(C_{1}x^{-\tau_{1}}+C_{2}x^{-\tau_{2}}+\ldots)F(x/x_{c})$
where $C_{i}$ are nonuniversal constants and that these corrections
do not determine universality class. The asymptotic scaling behavior
is determined by the leading power law. We assume that the deviation
from a simple power-law for densities $0<c\leq0.1$ (Sec. \ref{sec:Results})
could be explained by this correction. We observed that the exponents
for large avalanches $a$ are larger than the approximate exponents
($\tau_{a,L=1024}=1.23$ in the Fig. \ref{cap:avalanche-area}) which
cover the whole range. As a consequence, the leading exponents $\tau_{a,L}(c)$
for densities $0<c<0.1$ are higher than the approximate exponents
which we found (they are not shown in the Fig.\ref{cap:avalanche-area}
for $0<c\leq0.09$). It is evident that the leading scaling exponents
$\tau_{a,L}$ are different and are not constant (Fig. \ref{cap:exponents})
as in the case of the BTW model or the Manna model and thus the model
for these densities belongs to a different class than the BTW model
or the Manna model.

Divergences from the expected power-law behaviour of the BTW model
and a need of sub-dominant correction were observed in another inhomogeneous
sandpile model \cite{Cer}. Here the avalanche dynamic was disturbed
by sites which had the second higher threshold. The effect was significant
for thresholds $E_{C}\geq32$ and low concentration of such sites
\cite{Cer}. 

The multifractal properties (Fig. \ref{cap:multi}) of the model given
by Eqs. (\ref{eq:thresholds})-(\ref{eq:relax4}) for the density
$c=0$ (the BTW model), and FSS for the density $c=1$ (the Manna
mode) agree well with the recent results \cite{Stella_2001}. In addition,
the crossover from multifractal to FSS was observed in the Fig. \ref{cap:-crossover}.
Our results can only predict that a critical density is expected to
be found in the interval of densities $0<c<0.01$ (Figs. \ref{cap:multi}
and \ref{cap:-crossover}). This interval is five times smaller than
what was found in Ref. \cite{Karma} where the results are based on
the autocorrelation function of the avalanche wave time series \cite{Menech_2}. 

We assume that divergences from power-law dependences in inhomogeneous
conservative models, \cite{Cer} and Eqs. (\ref{eq:thresholds})-(\ref{eq:relax4}),
have a common reason which is connected to the crossover from multifractal
scaling to FSS \cite{Karma}. In both models a disorder is induced
by deployment of disturbing sites. These disturbing sites either increase
the short range coupling during relaxations in deterministic model
\cite{Cer} or introduce the random toppling [Eq. (\ref{eq:relax4})].
In these models toppling imbalance \cite{Karma_E,Karma} only for
a few such sites can change character of waves in the models from
coherent to more fragmented waves \cite{Ben-Hur,Biham,Milsh,Stella_2001}.

In this study, the multifractal properties of the BTW model which
is initially homogeneous, are destroyed at very low concentrations
of such disturbing sites. In the opposite case, the Manna model shows
the FSS and resistance to disturbance caused by presence of BTW sites
because all significant exponents from Eq. (\ref{eq:FSS}) are approximately
constant in a broad range of densities $0.15\leq c\leq1$. One possible
explanation for this is that the nature of the small perturbation
of the model is not the same when we perform changes around the densities
at $c=0$ and $c=1$. A small perturbation of the dynamical rules
of the BTW model ($c=0$) breaks the toppling symmetry \cite{Karma}
and this may explain why the changes in the scaling exponents $\tau_{x}(c)$
and capacity fractal dimension $D_{s}(c)$ are so unexpected. On the
other hand, for the Manna model ($c=1$), decreasing of the density
$c$ cannot influence the unbalanced toppling symmetry of the Manna
model \cite{Karma}. For sandpile models which show FSS this is an
expected result and agrees well with the theory \cite{Pietronero,Chessa},
where a small modification of toppling rules cannot change the scaling
exponents. 

We can clearly identify two universality classes which correspond
to the classes proposed in papers \cite{Ben-Hur} or \cite{Karma}:
(a) nondirected models, for density $c=0$ (BTW model, the multifractal
scaling \cite{Tebaldi_1999,Stella_2001,Menech}), and they show a
precise toppling balance \cite{Karma} and they are sensitive on disturbance
of avalanche dynamics, (b) random relaxation models, for densities
$0.1<c<1$ where FSS of $P(x)$ is verified, they are nondirected
only on average (Manna two-state model $c=1$ \cite{Ben-Hur}). In
these models breaking of the precise toppling balance \cite{Karma}
is observed, the scaling exponents are resistant to disturbance of
avalanches. The classification for densities $0<c<0.1$ is not so
clear. If we follow the proposed classifications then the model is
a random relaxation model \cite{Ben-Hur} with broken precise toppling
balance \cite{Karma} and it belongs in the same class as the Manna
model. On the other hand, the scaling exponents differ from the Manna
model and they are not universal ($\tau_{x}(c)\neq const$., $D_{s}(c)\neq const.$),
and the reasons of the sub-dominant approximation of area probability
distribution functions \cite{Chessa} can play an important role.
We assume that a new universality class between the BTW ($c=0$, multifractal
scaling) and the Manna ($c>0.5$, FSS) classes \cite{Karma,Karma_E}
could be identified for densities $0<c<0.1$. However, a more detailed
study is necessary to verify this classification. 

Our additional arguments to the previous results \cite{Ben-Hur,Biham,Cer,Karma,Milsh,Menech,Lubeck}
show that small modifications of the dynamical rules of the model
can lead to different universality classes what is considered to be
unusual from a theoretical standpoint \cite{Chessa}.

\section{\label{sec:Conclusion}Conclusion}

In these computer simulations multifractal scaling of the BTW model
\cite{Tebaldi_1999} and FSS of the Manna model \cite{Stella_2001}
were confirmed. In addition, a crossover from multifractal scaling
to FSS \cite{Karma} was observed when avalanche dynamics of the BTW
model was disturbed by Manna sites which were randomly deployed in
the lattice, as their density was increased. This crossover takes
place for a certain density $c$ in the interval $0<c<0.01$. This
interval is five times smaller than what was found recently \cite{Karma}.
The scaling exponents $\tau_{x}(c)$ and the capacity fractal dimension
$D_{s}(c)$ are not constant for all densities $c$ which is necessary
if the models \cite{BTW_1987,Manna_1991} belong to the same universality
class. These result agree well with the previous conclusions that
multifractal properties of the BTW model \cite{Stella_2001,Tebaldi_1999,Menech},
toppling wave character \cite{Ben-Hur,Biham,Milsh} and precise toppling
balance \cite{Karma,Karma_E} are important properties for solving
the universality issues.

An open question remains about how to characterize the universality
class for densities $0.01<c<0.1$, where the scaling exponents are
not universal ($\tau_{x}(c)\neq const$. and $D_{s}(c)$$\neq const.$)
and in addition, the avalanche probability distributions $P(a)$ do
not show exact power-law behavior since the sub-dominant corrections
of $P(a)$ \cite{Chessa} are important. In this interval of densities
$c$, our model belongs to the random relaxation models \cite{Ben-Hur}
and to the models with unbalanced toppling sites \cite{Karma,Karma_E},
however, its scaling exponents are not equal to the exponents of the
Manna model.

Based on the previous findings \cite{Karma,Karma_E} and our results
we assume that the avalanche dynamics of undirected conservative models,
in which some of the probability distribution functions show a multifractal
scaling (the BTW model), is disturbed by suitable toppling rules which
are different from the two-state Manna model (for example a stochastic
four-state Manna model \cite{Ben-Hur,Milsh}), then a local manner
for the energy distribution during the relaxation can be important
and can change the scaling exponents. However, the models which show
the FSS for all probability distribution functions (the Manna model)
are not sensitive to the details of the toppling rules and are consistent
with theoretical predictions \cite{Pietronero,Chessa}.

\section*{Acknowledgments}

The author thanks G. Helgesen for his comments to the manuscript.
The numerical simulations were carried out using the ARC middleware
and NorduGrid infrastructure \cite{Nordugrid}. We acknowledge the
financial support from the Slovak Ministry of Education: Grant NOR/SLOV2002.


\begin{thebibliography}{10}
\bibitem{BTW_1987}P. Bak, C. Tang, and K. Wiesenfeld, Phys. Rev. Lett. \textbf{59},
381 (1987); Phys. Rev. A \textbf{38}, 364 (1988).
\bibitem{Manna_1991}S. S. Manna, J. Phys. A \textbf{24}, L363 (1991).
\bibitem{Kadanoff}L. P. Kadanoff, S. R. Nagel, L. Wu and S. Zhou, Phys. Rev. A \textbf{39},
6524 (1989). 
\bibitem{Pietronero}L. Pietronero, A. Vespignani, and S. Zapperi, Phys. Rev. Lett. \textbf{72},
1690 (1994); A. Vespignani, S. Zapperi, and L. Pietronero, Phys. Rev.
\textbf{}E, \textbf{51}, 1711 (1995).
\bibitem{Menech}M. De Menech, Phys. Rev. E \textbf{70}, \textbf{}028101 (2004).
\bibitem{Lubeck}S. L\"{u}beck and K. D. Usadel, Phys. Rev. E \textbf{55}, 4095 (1997).
\bibitem{Ben-Hur}A. Ben-Hur and O. Biham, Phys. Rev. E \textbf{53}, R1317 (1996).
\bibitem{Biham}E. Milshtein, O. Biham, and S. Solomon, Phys. Rev. E \textbf{58},
303 (1998).
\bibitem{Milsh}O. Biham, E. Milshtein, and O. Malcai, Phys. Rev. E \textbf{63}, 061309
(2001).
\bibitem{Chessa}A. Chessa, H. E. Stanley, A. Vespignani, and S. Zapperi, Phys. Rev.
E \textbf{59}, \textbf{}R12 (1999).
\bibitem{Tebaldi_1999}C. Tebaldi, M. De Menech and A. L. Stella, Phys. Rev. Lett. \textbf{83},
3952 (1999).
\bibitem{Stella_2001}A. L. Stella and M. De Menech, Physica A \textbf{295}, 101 (2001).
\bibitem{Karma}R. Karmakar, S. S. Manna, and A. L. Stella, Phys. Rev. Lett. \textbf{94},
088002 (2005).
\bibitem{Karma_E}R. Karmakar and S. S. Manna, Phys. Rev. E \textbf{71}, 015101 (2005). 
\bibitem{Cer}J. \v{C}ern\'{a}k, Phys. Rev. E \textbf{65}, 046141 (2002).
\bibitem{Dhar}D. Dhar, Physica A \textbf{263}, 4 (1999).
\bibitem{Manna_PA}S. S. Manna, Physica A \textbf{179}, 249 \textbf{}(1991).
\bibitem{Lubeck_2}S. L\"{u}beck, Phys. Rev. E \textbf{61}, 204 (2000). 
\bibitem{Christensen}K. Christensen, H. C. Fogedby, and H. J. Jensen, J. Stat. Phys. \textbf{63},
653 (1991).
\bibitem{Menech_2}M. De Menech and A. L. Stella, Phys. Rev. E \textbf{62}, R4528 (2000). 
\bibitem{Nordugrid}For more information see the web site\\ http://www.nordugrid.org 
\end{thebibliography}
\end{document}